\documentclass[pra,showpacs,floatfix,superscriptaddress,aps,twocolumn]{revtex4}
\usepackage{amsmath}
\usepackage{amssymb}

\usepackage{graphicx}
\usepackage[usenames]{color}

\newcommand{\jpb}{J. Phys. B\ }

\begin{document}

\title{Dipolar Bose-Einstein condensate in a ring or in a  shell}

\author{  S. K. Adhikari\footnote{Email: adhikari@ift.unesp.br, URL: http://www.ift.unesp.br/users/adhikari/}}
\affiliation{Instituto de F\'{\i}sica Te\'orica, UNESP - Universidade Estadual Paulista, 01.140-070 S\~ao Paulo, S\~ao Paulo, Brazil}

\begin{abstract}

We study properties of a trapped dipolar Bose-Einstein condensate (BEC) in a circular ring or a spherical shell using the 
mean-field Gross-Pitaevskii equation.  In the case of the ring-shaped trap we consider different orientations of the ring with respect to
the polarization direction of the dipoles. 
In the presence of long-range anisotropic dipolar and short-range contact interactions, 
the anisotropic density distribution of the dipolar BEC in both traps is discussed in detail. The stability condition of the dipolar 
BEC in both traps is illustrated in phase plot of dipolar and contact interactions. We also study and discuss the properties of a vortex 
dipolar BEC in these traps.

 \end{abstract}


\maketitle

\section{Introduction}

After the experimental realization of  
dilute trapped Bose-Einstein condensate (BEC) of alkali-metal atoms \cite{rmp}, it was realized that for attractive
atomic contact interaction the BEC is stable for the atomic interaction (measured by the scattering length)  
and number of atoms below a certain limit depending on the 
trap \cite{rmp,hulet,gammal}. In the case of $^7$Li atoms with attractive atomic interaction, in the harmonic trap used in  experiment \cite{hulet}, the condensate was stable for less than 
about 1400 atoms.   For repulsive contact interaction, the trapped BEC is unconditionally stable  for all 
values scattering length and number of atoms \cite{rmp}.

More recently, there has been experimental observation of BECs of $^{52}$Cr \cite{cr1,cr2}, $^{164}$Dy \cite{dy1,dy2}, and 
$^{168}$Er \cite{er}   with  large 
long-range anisotropic magnetic dipolar interaction. Bosonic polar molecules with much larger electric dipolar interaction 
are also being considered for BEC experiments
\cite{cr2,kkni}.
Thus one can study the properties of a
dipolar BEC  with variable short-range interaction \cite{cr1,dy2} using a Feshbach resonance \cite{fesh}. 
The dipolar BEC \cite{pla} with anisotropic long-range atomic interaction 
has many distinct features \cite{cr1,cr2,cr3,cr4,cr5}. The stability of a dipolar BEC depends on the scattering length
as well as  the trap geometry \cite{cr1,cr3,cr5}. A disk-shaped trap, with the polarization $z$ direction perpendicular to the plane of the disk, 
leads to  a repulsive dipolar interaction making the dipolar BEC 
more stable \cite{cr1}. On the other hand, a cigar-shaped dipolar BEC  oriented  along the polarization direction leads to 
an attractive dipolar interaction and hence may favor 
a collapse instability \cite{cr1,cr5,cr51,cr6}. Also, the anisotropic dipolar interaction leads, in general,  to a distinct anisotropic density distribution 
in a dipolar BEC. The shock and sound waves also propagate with different velocities in different directions in a dipolar BEC \cite{shock}.
Anisotropic collapse has been observed and studied in a dipolar BEC of $^{52}$Cr atoms \cite{dcol}.

The properties of a BEC have been studied on different types of traps, such as, the harmonic trap \cite{rmp}, optical-lattice (OL) trap \cite{olt}, 
bichromatic OL trap \cite{bich}, optical  speckle potential  trap \cite{spe},  double-well 
trap  \cite{dbl}, 
toroidal trap
\cite{tor,tor2},  ring-shaped trap in one \cite{1dr} and three \cite{3dr} dimensions,
among others. To study the properties
of the dipolar BEC, apart from the harmonic trap, the following traps have been used:
OL trap \cite{dol}, bichromatic OL trap \cite{dbol}, toroidal trap \cite{dt} and double-well trap \cite{ddbl}.

Very recently, there has been experimental realization of ring-shaped and 
spherical-shell-shaped traps \cite{rs1,rs2}. 
The ring geometry was created
by the time-averaged adiabatic potential resulting from the application of an oscillating magnetic bias field to a
radio-frequency-dressed quadrupole trap \cite{rs2}. The shell geometry was made from a
cylindrically symmetric quadrupole field with its symmetry
axis aligned with gravity \cite{rs2}.
These geometries of the trap present 
an opportunity to study the superfluid properties of a BEC
in a multiply connected geometry \cite{rs2} not found in usual traps. 
In this paper, we study the properties of a dipolar BEC in ring- and shell-shaped  traps.
The shell-shaped trap   is spherically symmetric, whereas the ring-shaped trap is at most axially symmetric.  
Different orientations of the ring-shaped dipolar BEC with respect to the polarization direction are considered. The stability of the dipolar 
BEC in these traps is illustrated  by  phase plots of dipolar and contact interactions. The anisotropic nature of stability properties in a
ring-shaped trap is further demonstrated by a consideration of the chemical potential of the system. The ring-shaped dipolar BEC is more stable 
when the plane of the ring is aligned perpendicular to the polarization direction, than when it is aligned parallel to the polarization 
direction.  In the former case the dipolar interaction is mostly repulsive and in the latter case it is mostly attractive. 
The anisotropic density distribution of the dipolar BEC in these traps is explicitly demonstrated. 
 
We also consider a vortex dipolar BEC (rotating around the polarization $z$ direction) of unit angular momentum \cite{vortex2,vortex}
in  ring- and shell-shaped traps, when the trapping potential is axially symmetric around the 
$z$ direction. In the case of a ring-shaped trap the axial symmetry is maintained  
when the ring is in a plane perpendicular to the $z$ direction. 
The ring-shaped 
vortex dipolar BEC is found to be nearly identical to the normal (nonrotating) dipolar BEC for the same sets of parameters.  The shell-shaped 
vortex dipolar BEC is found to possess a distinct density distribution when compared with a normal dipolar BEC.

In Sec. II we describe the ring- and shell-shaped traps and 
present the  mean-field Gross-Pitaevskii (GP) equation which we use  to study the normal and vortex 
dipolar BEC in these traps. In Sec. III we present the numerical results obtained by solving the 
GP equation using the Crank-Nicolson approach. We present  stability phase plots of the dipolar BECs in 
terms of contact and dipolar interactions. The anisotropic density distribution of the BECs, which is a 
consequence of the anisotropic dipolar interaction, is illustrated for both the traps. The difference 
in the density distribution between the normal and vortex BECs is also demonstrated.    
Finally in Sec. IV we present a brief summary and concluding remarks.

\section{Analytical  Consideration}


A trap in the shape of a  spherical shell will be taken in the form
\begin{eqnarray}
V({\bf r})= V_0 \times \frac{1}{2}m\omega^2
 (r-r_0)^2,
\end{eqnarray}
with ${\bf r}=\{x,y,z\}, r=\sqrt{(x^2+y^2+z^2)}$, where $m$ is the mass of an atom, $r_0$ is the radius of the shell, 
$V_0$ is the strength of the trap and $\omega$ the
frequency.
For $r_0=0$,  trap (\ref{shellpot}) reduces to the usual harmonic trap.
 Similarly, the ring-shaped trap can be written as  
\begin{eqnarray} 
V({\bf r})=V_0  \times \frac{1}{2}m\omega^2
[  (\sqrt{x^2+p^2}-r_0)^2 +q^2],
\end{eqnarray}
where $p=(y\cos\alpha+z\sin\alpha), q= (z\cos\alpha-y\sin\alpha),$ 
$V_0$ is the strength of the trap, 
$\alpha$ is the angle between the polarization direction $z$ 
and the perpendicular to the plane of the ring, 
and $r_0$ is the radius of the ring. For $r_0=0$ and $\alpha=0$ or $\pi/2$,  trap 
(\ref{ringpot}) reduces to the harmonic trap.

A dilute
 dipolar BEC  of $N$ atoms, in shell- or ring-shaped traps (\ref{shellpot}) or (\ref{ringpot}) will be studied
 using the  following Gross-Pitaevskii  (GP) 
equation  \cite{cr1,cr3}
\begin{eqnarray}  \label{gp3d} 
i \frac{\partial \phi({\bf r},t)}{\partial t}
& =&  \biggr[ -\frac{\nabla^2}{2} + V  +g|\phi|^2+g_{dd}
F\biggr] \phi({\bf r},t),
\end{eqnarray} 
with    $g=4\pi a N, g_{dd}=3a_{dd}N$.
Here the dipolar nonlinearity 
$ F= \int U_{dd}({\bf r -r'})|\phi({\bf r'},t)|^2
d{\bf r'}$,  
$ U_{dd}({\bf R}) =  
(1-3\cos^2\theta)/R^3 $,
 ${\bf R=r-r'},$  
 normalization $\int \phi({\bf r})^2 d {\bf r}$ = 1,
    $\theta$ 
the angle between $\bf R$ and the polarization direction   $z$.  For the shell-trap
\begin{equation} \label{shellpot}
V\equiv  V_{\mbox{shell}}=V_0 
 (r-r_0)^2,\end{equation}
 and for the ring-trap 
\begin{equation}\label{ringpot}
V \equiv V_{\mbox{ring}}=V_0 [  (\sqrt{x^2+p^2}-r_0)^2 +q^2],
\end{equation} 
$a_{dd}
=\mu_0	\tilde \mu^2 m /(12\pi \hbar^2)$ 
the strength of 
dipolar interaction,  $\mu_0$ 
the permeability of free space, and
 $	\tilde \mu$ the (magnetic) dipole moment. 
In  Eq. (\ref{gp3d}), 
length is measured in 
units of  $l_0 \equiv  \sqrt{\hbar/m\omega}$, potential and energies in units of $\hbar\omega$, 
 time $t$ in units of $t_0 = \omega^{-1}$. In this work we conveniently take $l_0=1$ $\mu$m, 
which is the typical unit of length in BEC experiments.

The ring-shaped trap with $\alpha =0$ and the 
shell-shaped  trap are axially symmetric around the $z$ axis and a  vortex BEC rotating 
around $z$ axis with a conserved angular momentum 
can be conveniently introduced in these cases. 
To obtain a quantized vortex of unit angular momentum $\hbar$
around $z$ axis,  we
introduce a phase (equal to the azimuthal angle) in the 
wave function \cite{vortex2}. This
procedure introduces a centrifugal term $1/[2(x^2+y^2)]$ in the potential of the 
GP equation   so that 
\begin{equation}\label{potv}
V= {\cal V} +\frac{1}{2(x^2+y^2)},
\end{equation}
where ${\cal V}$ is $ V_{\mbox{shell}}$ for the shell-shaped trap and $ V_{\mbox{ring}}$ for ring-shaped trap
with $\alpha=0$. For the ring-shaped trap with $\alpha \ne 0$, the potential is not axially symmetric 
and a conserved angular momentum cannot be defined. 
We
adopt this procedure to study an axially-symmetric vortex in a ring- and shell-shaped  dipolar
BEC. 

The chemical potential $\mu$  of a stationary state propagating as $\phi(t)\sim \exp(-i\mu t)$
is defined by 
\begin{equation}
\mu = \int \biggr[ -\frac{1}{2}|\nabla \phi|^2 + V|\phi|^2  +g|\phi|^4+g_{dd}
F|\phi|^2\biggr]d{\bf r}. 
 \end{equation}

\section{Numerical Calculation}

We numerically solve  the 3D GP equation (\ref{gp3d})
using   the split-step  Crank-Nicolson 
method  \cite{Muruganandam2009}.  The dipolar potential is divergent at short
distances and hence the treatment of this potential requires some care.
The integral over the dipolar potential is evaluated in Fourier (momentum) space by a convolution 
identity \cite{cr3} requiring the Fourier transformation of the dipolar potential and 
density. The Fourier transformation of the dipolar potential can be analytically evaluated  
\cite{cr3}. The remaining Fourier transformations are evaluated numerically using 
a fast Fourier transformation algorithm. In the Crank-Nicolson discretization we used  
space step 0.1,  time step 0.002 and up to  256 space discretization points in each of the 
three Cartesian directions. We performed an analysis of errors of sizes and energies calculated 
using our routine and find that the maximum numerical error in the calculation 
is less than 0.5 $\%$.

\begin{figure}[!t]
\begin{center}
\includegraphics[width=\linewidth,clip]{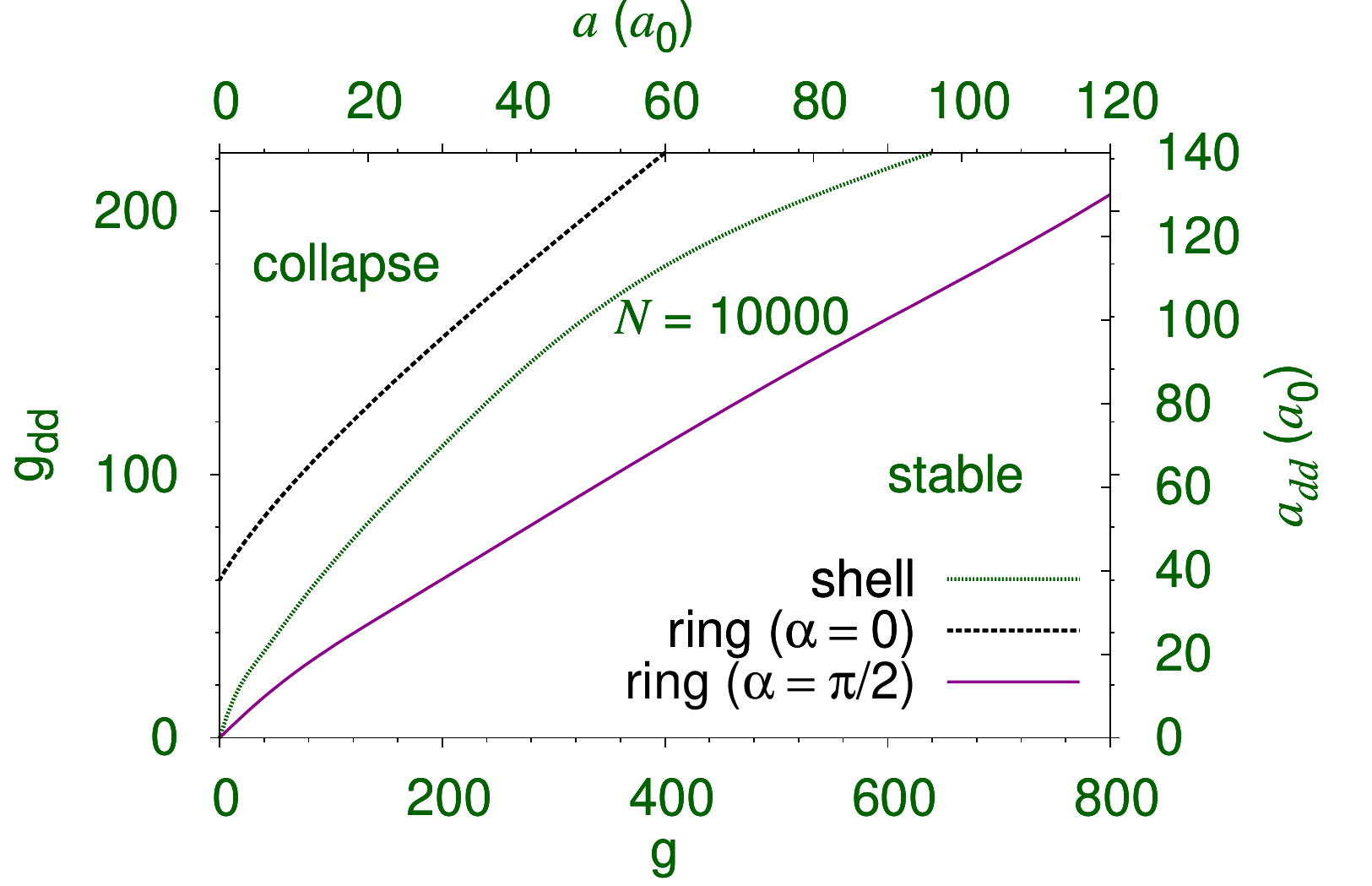}
\end{center}

\caption{(Color online)
Phase plot of contact and dipolar nonlinearities $g (=4\pi a N)$ and 
$g_{dd} (=3a_{dd}N)$ showing the stability 
line for 
shell- and ring-shaped dipolar BECs for $\alpha=0$ and $\pi/2$.
The BEC is stable in the region below these lines and unstable above.   
 }
\label{fig1}
\end{figure}

\subsection{Normal dipolar BEC}

In Eqs. (\ref{shellpot}) and (\ref{ringpot}), for shell and ring shapes, we take the strength 
of the potentials $V_0=10$ and radius $r_0=10$. This makes  reasonably strong traps, so that 
the widths of the shell or ring is small compared to the radius of the shell 
or ring and the  shell  and ring shapes 
of the BECs are pronounced. 
First we study the stability of the shell- and ring-shaped dipolar BEC.  The stability properties 
of these BECs are best illustrated in phase plots involving the strengths of 
contact and dipolar nonlinearities $g$ and $g_{dd}$ as shown in Fig. \ref{fig1}. 
For the ring shape, we show two orientations of the ring for $\alpha=\pi/2$ and $0$, the 
former corresponding to a ring in a plane parallel to the polarization direction $z$ and the latter 
in a plane perpendicular to the polarization direction $z$. The orientation of the dipoles in the former
is quite similar to that in a cigar-shaped BEC along $z$ axis in a harmonic trap, and that of the latter is similar to a 
disk-shaped BEC in $x-y$ plane
in a harmonic trap. The dipolar interaction contributes predominantly 
attractively in a cigar configuration with dipoles 
arranged parallel to the length of the cigar and it contributes predominantly 
repulsively in a  disk configuration with 
dipoles arranged perpendicular to the plane of the disk. Hence for a ring-shaped dipolar BEC with $\alpha =0$
the dipolar interaction contributes repulsively and to a positive quantity in energy and chemical potential. 
On the other hand,   for a ring-shaped dipolar BEC with $\alpha =\pi/2$
the dipolar interaction contributes attractively and to a negative quantity in energy and chemical potential. 
Consequently, for $\alpha=0,$  the ring-shaped dipolar BEC is stable for a reasonably large dipolar interaction below a critical 
value ($g_{dd}<g_{dd}^{\mbox{crit}}$)
without any contact interaction ($g=0$).   However, the dipolar BEC without any contact interaction  ($g=0$) collapses
for larger dipolar interactions  ($g_{dd}>g_{dd}^{\mbox{crit}}$).
For $\alpha =\pi/2,$ the dipolar interaction is mostly attractive and for a 
non-zero dipolar interaction,
the ring-shaped dipolar BEC is stable only for 
contact interaction above a critical value as shown by the phase plot in Fig. \ref{fig1}.

Next  we consider the shell-shaped dipolar BEC. In this case the trapping potential is spherically symmetric and unless 
the dipolar interaction is large compared to the contact interaction, the contribution of the dipolar interaction to 
the stability of the BEC, and to the chemical potential of the system, is  insignificant. However,    
for a finite $g_{dd}$ the system collapses unless $g$ is larger 
than a critical value.   To avoid the collapse, a finite 
amount of contact repulsion is to be introduced through a finite $g$ as shown in Fig. \ref{fig1}.

\begin{figure}[!t]
\begin{center}
\includegraphics[width=\linewidth,clip]{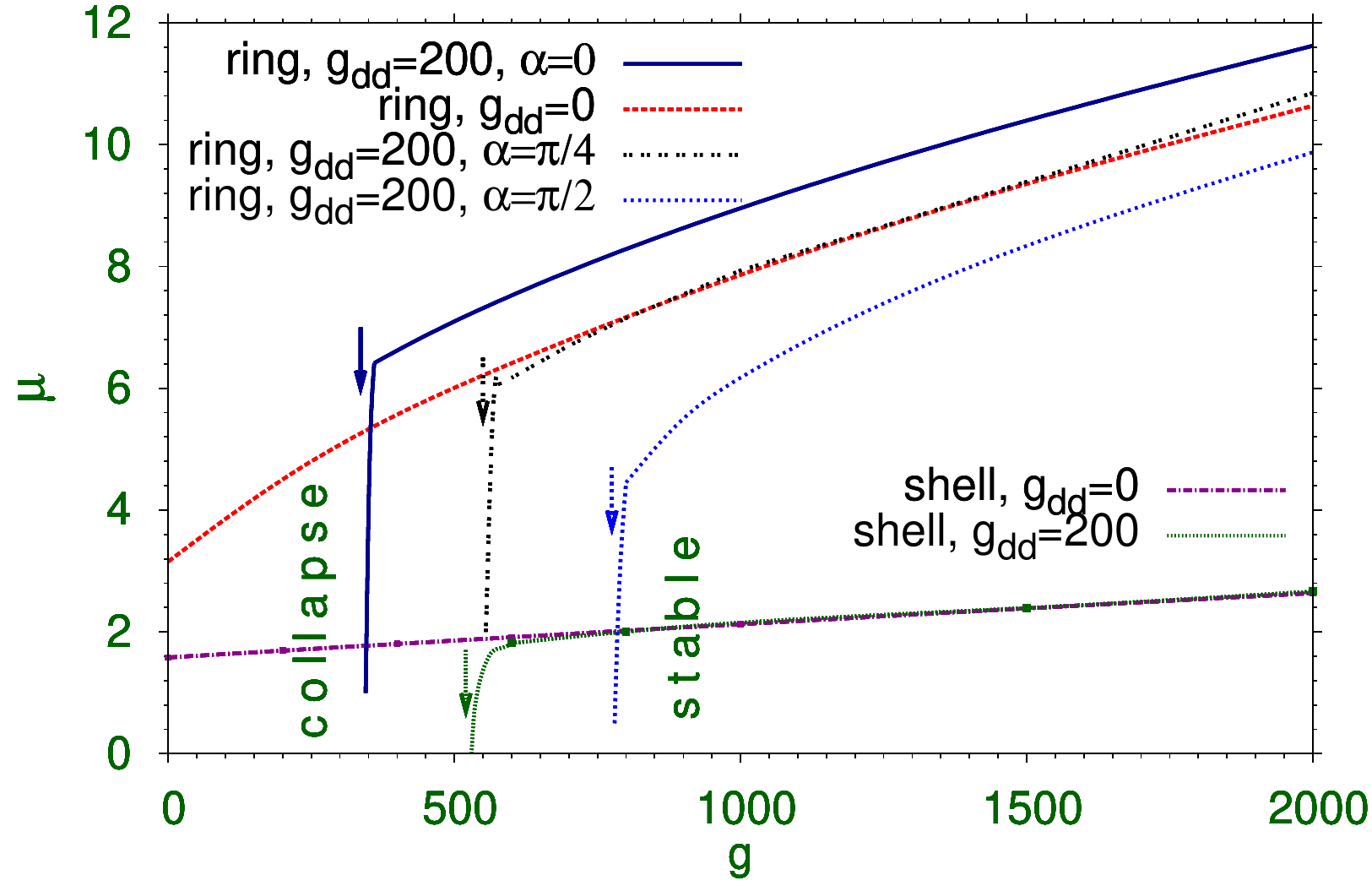}
\end{center}

\caption{(Color online) The chemical potential $\mu$ versus the contact-interaction  nonlinearity $g$ for the ring- and shell-shaped BEC for different dipolar nonlinearity $g_{dd}$. We show 
the results for $g_{dd}=0$ and 200 for $\alpha=0, \pi/4, $ and $\pi/2$.
 The arrow represents the onset of collapse in different cases. }
\label{fig2}
\end{figure}

Further consequence of stability characteristics  is shown in Fig. \ref{fig2}, where 
we plot the chemical potential of the dipolar BEC versus $g$ for different values of $g_{dd} (=0, 200)$. For a ring-shaped 
dipolar BEC with $\alpha =\pi/2$, the dipolar interaction is attractive and contributes negatively 
to the chemical potential; whereas, for $\alpha=0$, the dipolar interaction is repulsive and contributes 
positively to the chemical potential. Consequently, the chemical potential for  $\alpha =0$
is the largest and for $\alpha=\pi/2$ is the smallest for a fixed $g$ and $g_{dd}$. The chemical potential 
for $\alpha=\pi/4$ lies close to the nondipolar BEC with $g_{dd}=0$, as the contribution of the dipolar 
term is small in this case. 
For a shell-shaped dipolar BEC, as noted before, the contribution of the dipolar 
interaction to the chemical potential is insignificant and the chemical potentials for $g_{dd}=0$ and 200 are  
practically the same.
{\color{red} The
dipolar interaction energy nearly 
vanishes when integrated over all angles in a spherically symmetric density
configuration \cite{cr51} as in the spherical  shell.
}
 However, in all cases of dipolar BEC $(g_{dd}=200)$, the system collapses for 
$g$ less than a critical value $g_c$ (indicated by an arrow in Fig. \ref{fig2}), where the chemical potential suddenly jumps to a infinitely large 
negative value corresponding to a collapsed state. 

\begin{figure}[!t]
\begin{center}
\includegraphics[width=.49\linewidth,clip]{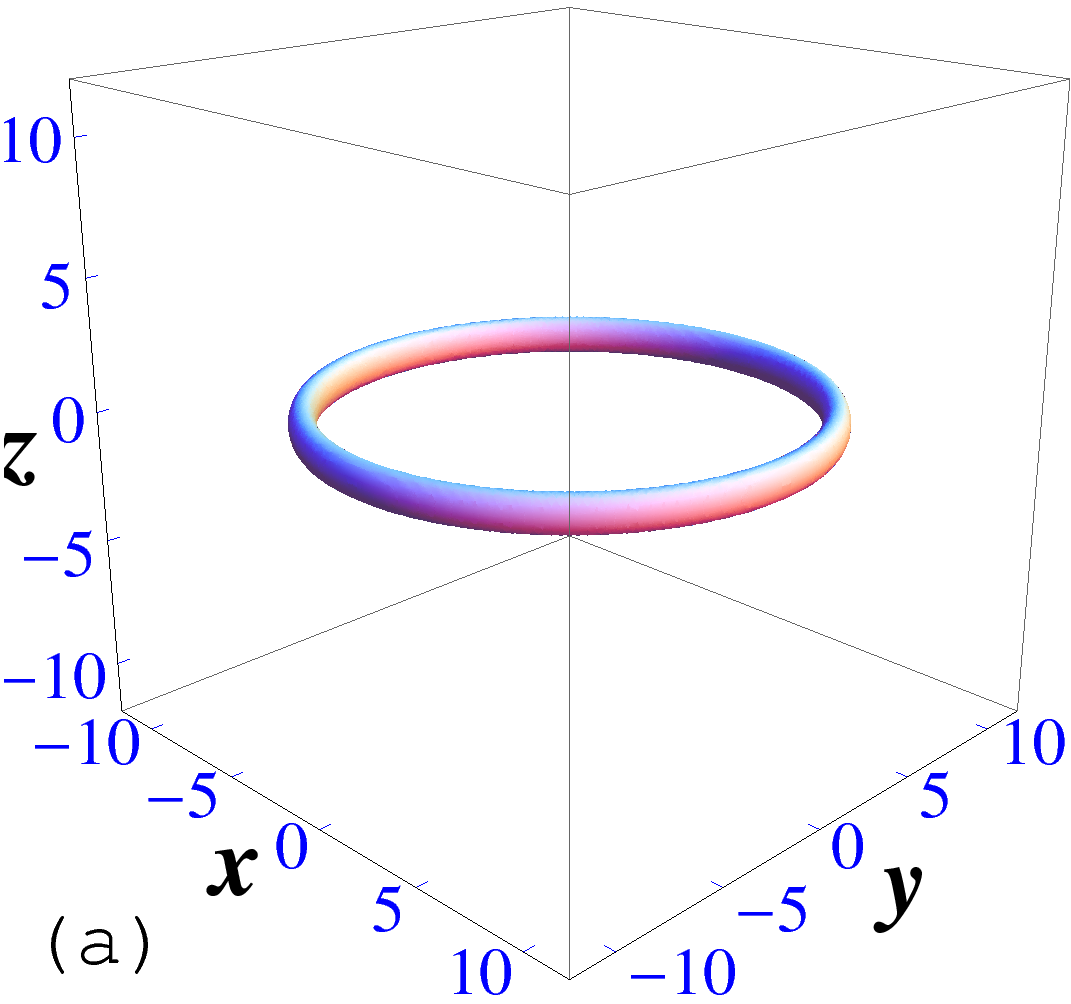}
\includegraphics[width=.49\linewidth,clip]{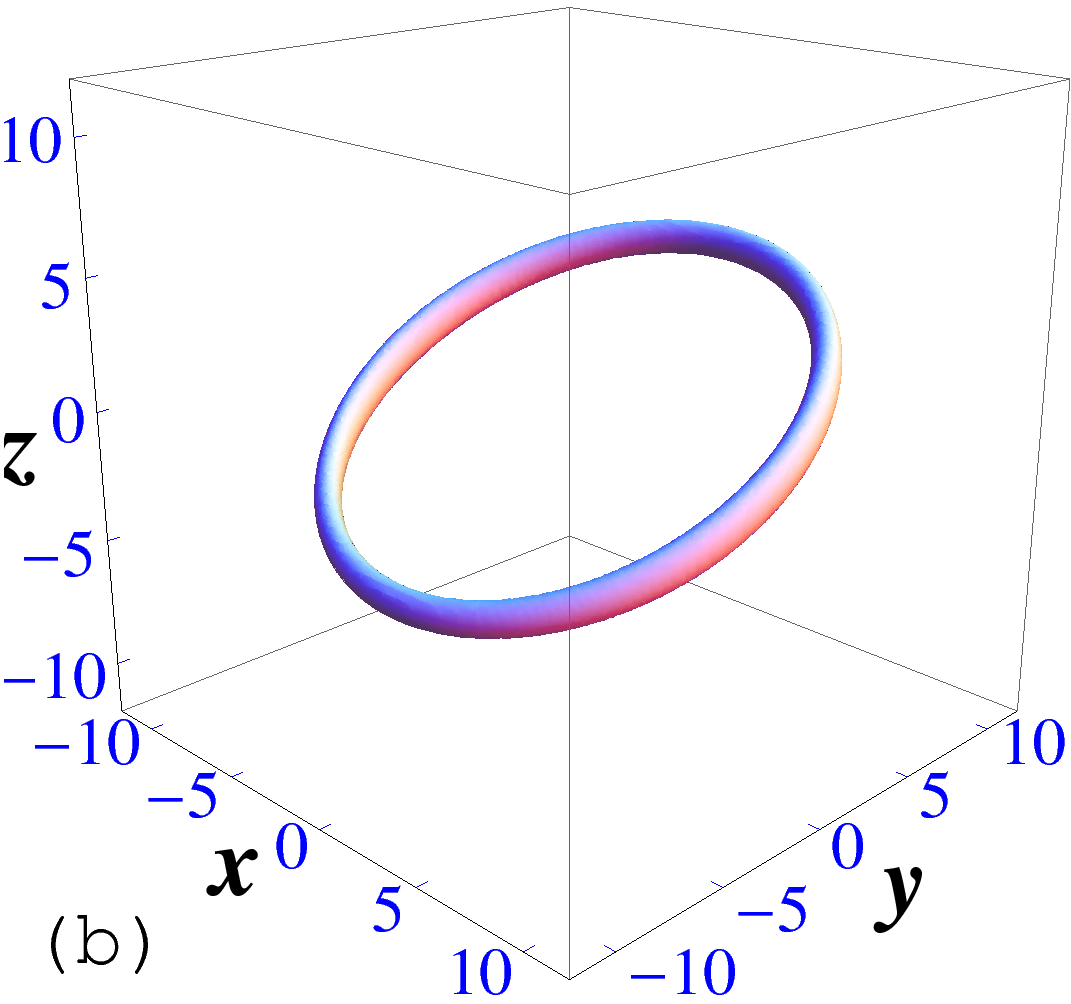}
\includegraphics[width=.49\linewidth,clip]{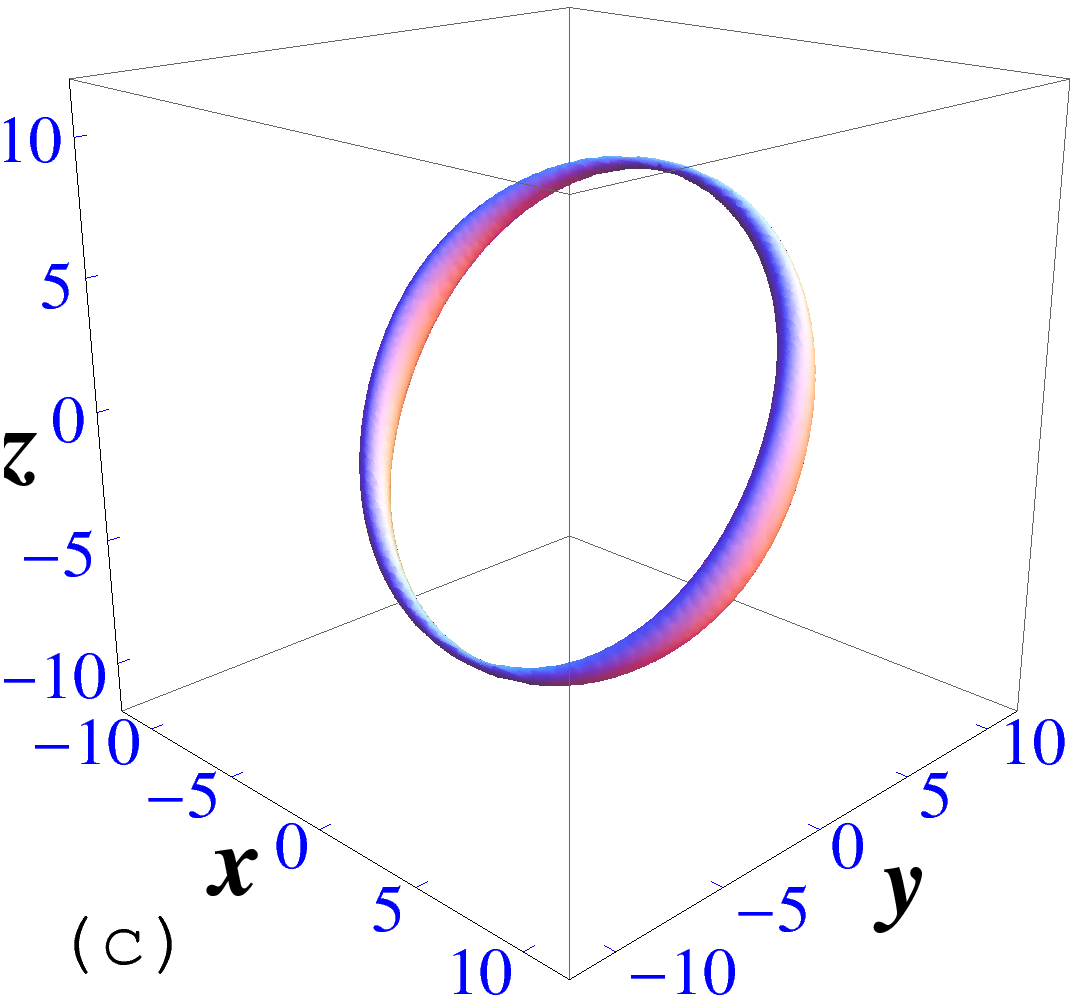}
\includegraphics[width=.49\linewidth,clip]{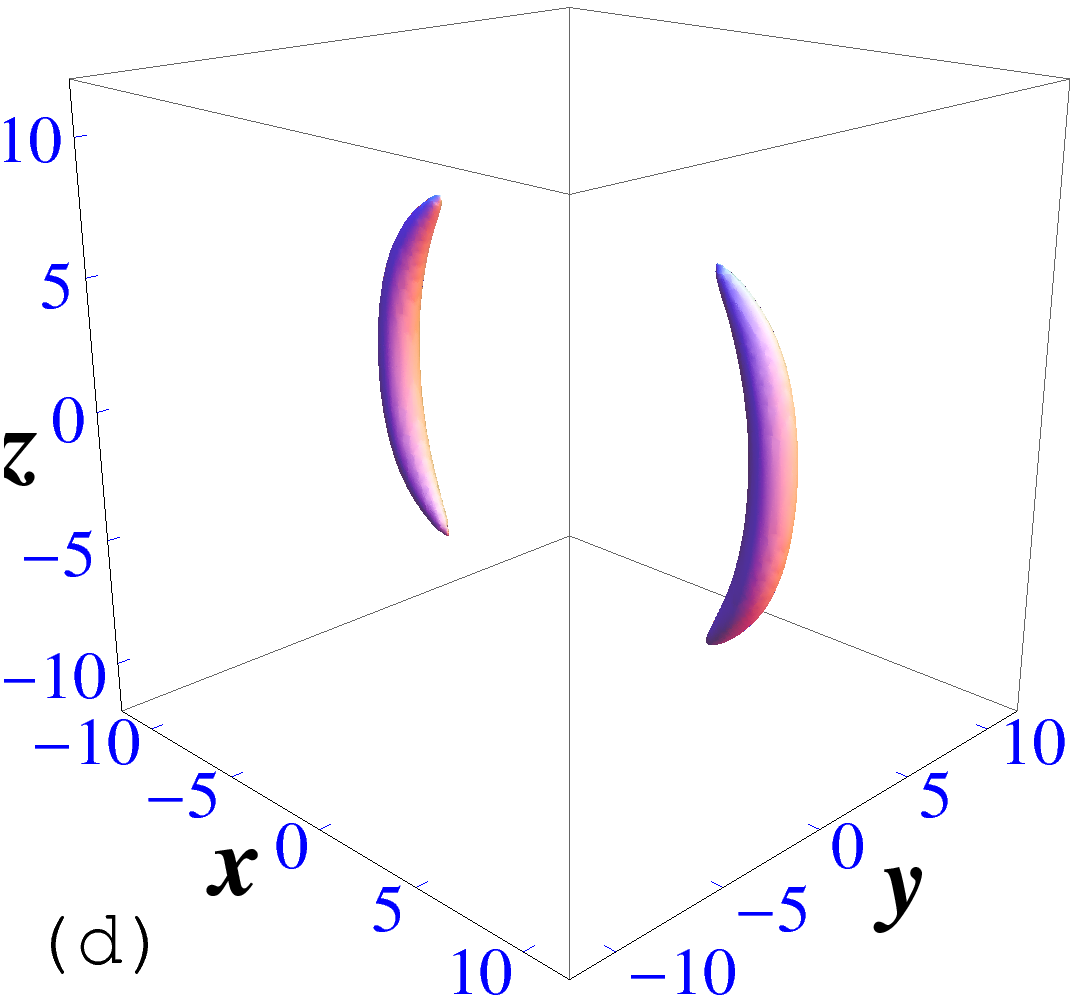}
\end{center}

\caption{(Color online) 
3D contour plot of density $|\phi|^2$ of a ring-shaped dipolar $^{164}$Dy  BEC
 for $ a_{dd}=130a_0, N=5000, a=120a_0 
$ and (a) $\alpha=0$, (b)$\alpha=\pi/6$, (c) $\alpha=\pi/3$, (d)  $\alpha=\pi/2$. The density at the 
contour is 0.005. {\color{red} The variables $x$, $y$, and $z$ are in units of $l_0
(= 1$ $\mu$m.)}
}  
\label{fig3}
\end{figure}

\begin{figure}[!b]
\begin{center}
\includegraphics[width=\linewidth,clip]{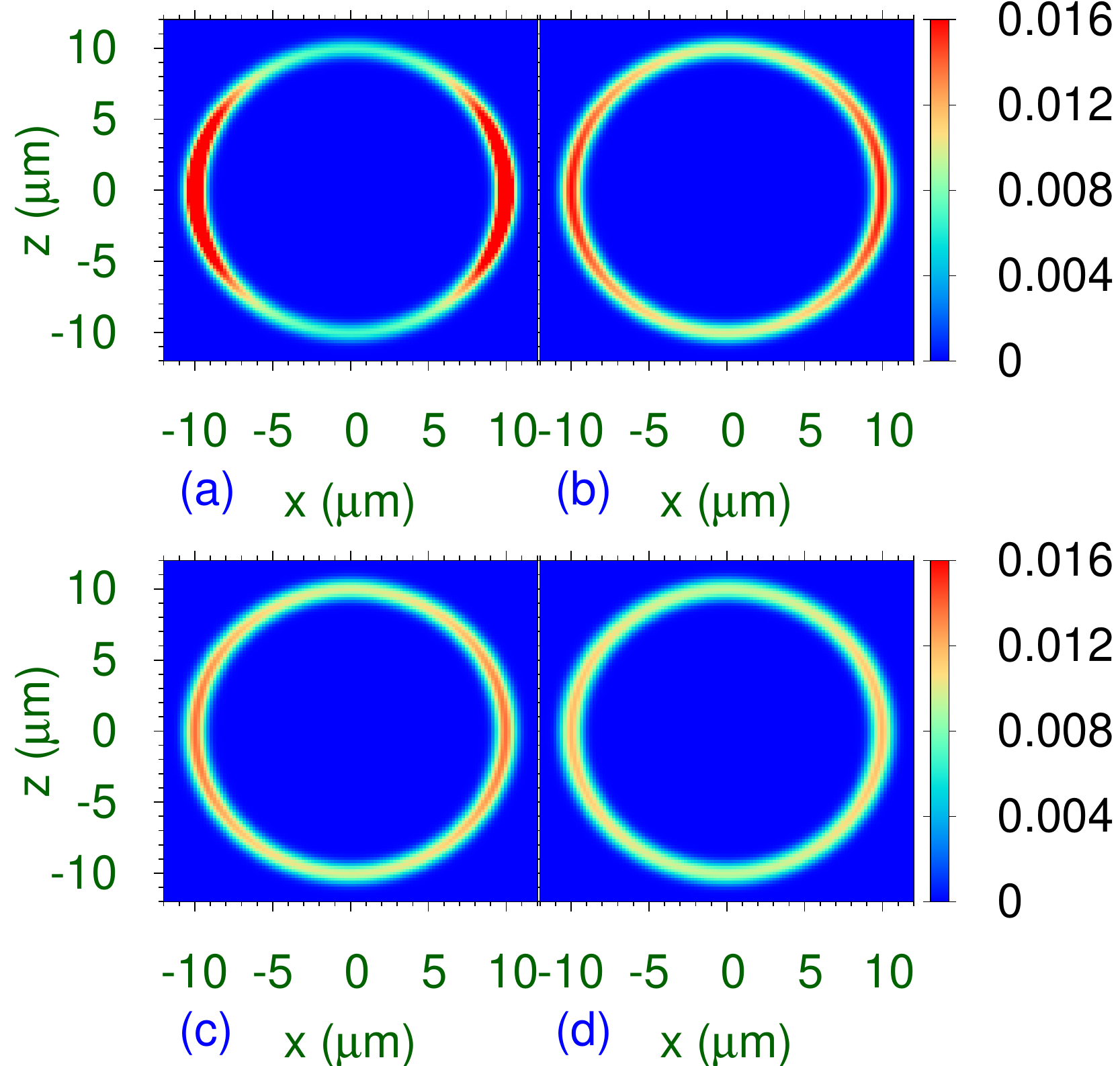}
\end{center}

\caption{(Color online) 
2D contour plot of density in the $x-z$ plane $|\phi(x, y=0,z)|^2$ 
for $\alpha = \pi/2$, for a ring-shaped dipolar BEC of 5000 $^{52}$Cr atoms 
with $a_{dd}=15a_0, 
$ and (a) $a=20a_0$, (b)$a=50a_0$, (c) $a=70a_0$, (d) $a=100a_0$. 
}  
\label{fig4}
\end{figure}

Next we illustrate the structure of a ring-shaped dipolar BEC by three-dimensional (3D) contour plots of density 
$|\phi|^2$. We exhibit the contour density plots of a $^{164}$Dy BEC of 5000 $^{164}$Dy atoms with  $a_{dd}=130a_0$ \cite{dy2}. The 
S-wave scattering length for contact interaction in this case is 
taken as $a=120a_0$ in close agreement with some experimental estimate \cite{dy2}. The 3D contour plots of density  
for $\alpha =0, \pi/6,\pi/3,$ and 
$\pi/2$ in this case  are shown in Figs. \ref{fig3} (a), (b), (c), and (d), respectively. For these parameters, in all the plots 
of Figs. \ref{fig3}, $g=399$ and $g_{dd}=103$. 
In the plane of the ring, the density is circularly symmetric for $\alpha 
=0$. However,  as the angle $\alpha$ increases due to the anisotropic dipolar interaction the density distribution in the plane of the ring
is no longer isotropic as can be seen in Figs. \ref{fig3} (b) and (c). The anisotropy in density distribution in the ring is visible 
for $\alpha = \pi/3$ in 
Fig. \ref{fig3} (c) and is
 most explicit for $\alpha=\pi/2$ in Fig. \ref{fig3} (d). 
{\color{red} Although, the dipolar interaction is long-range in nature, 
it is practically zero across the diameter of the ring 
or the shell (20 micron or 200,000 \AA).}

\begin{figure}[!t]
\begin{center}
\includegraphics[width=.49\linewidth,clip]{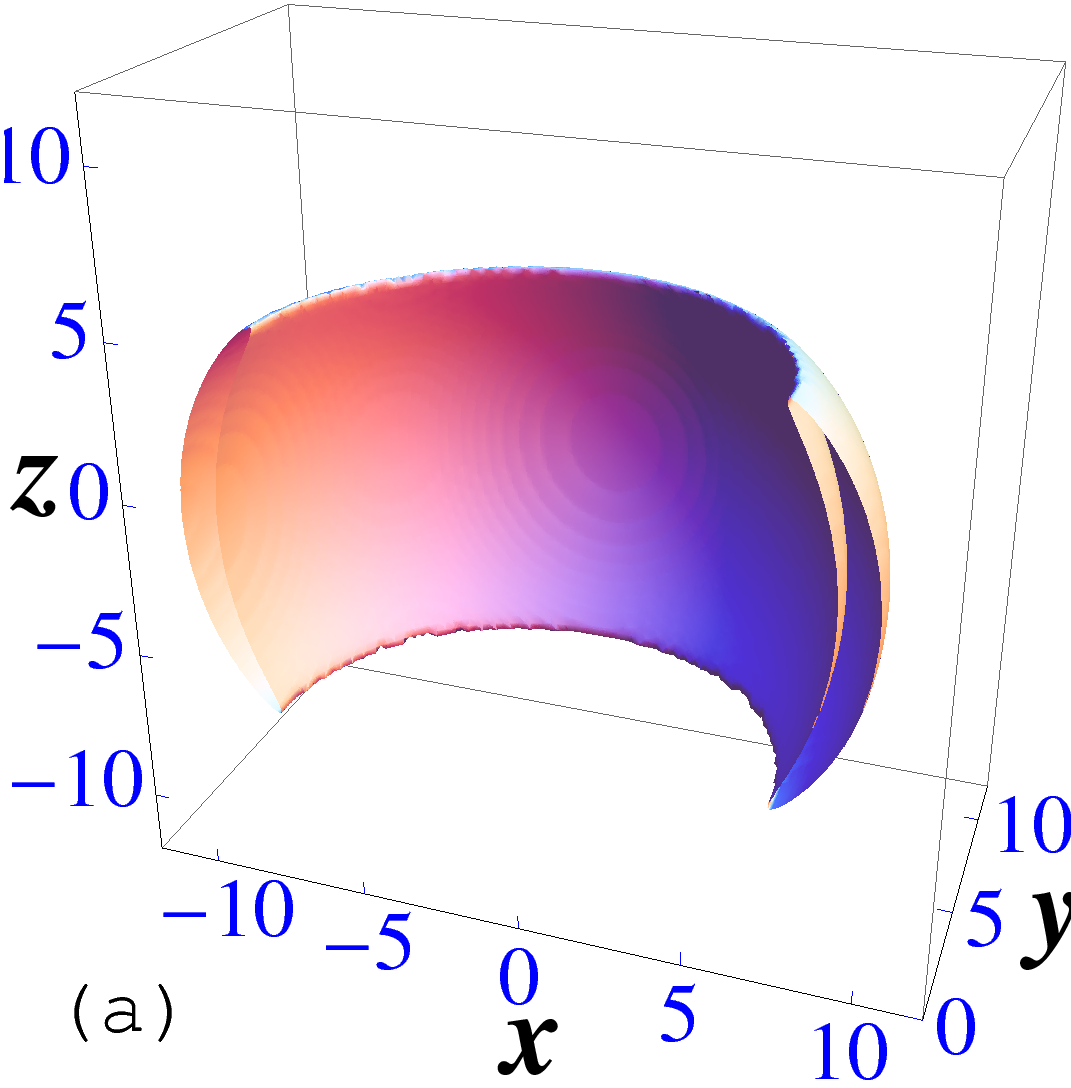}
\includegraphics[width=.49\linewidth,clip]{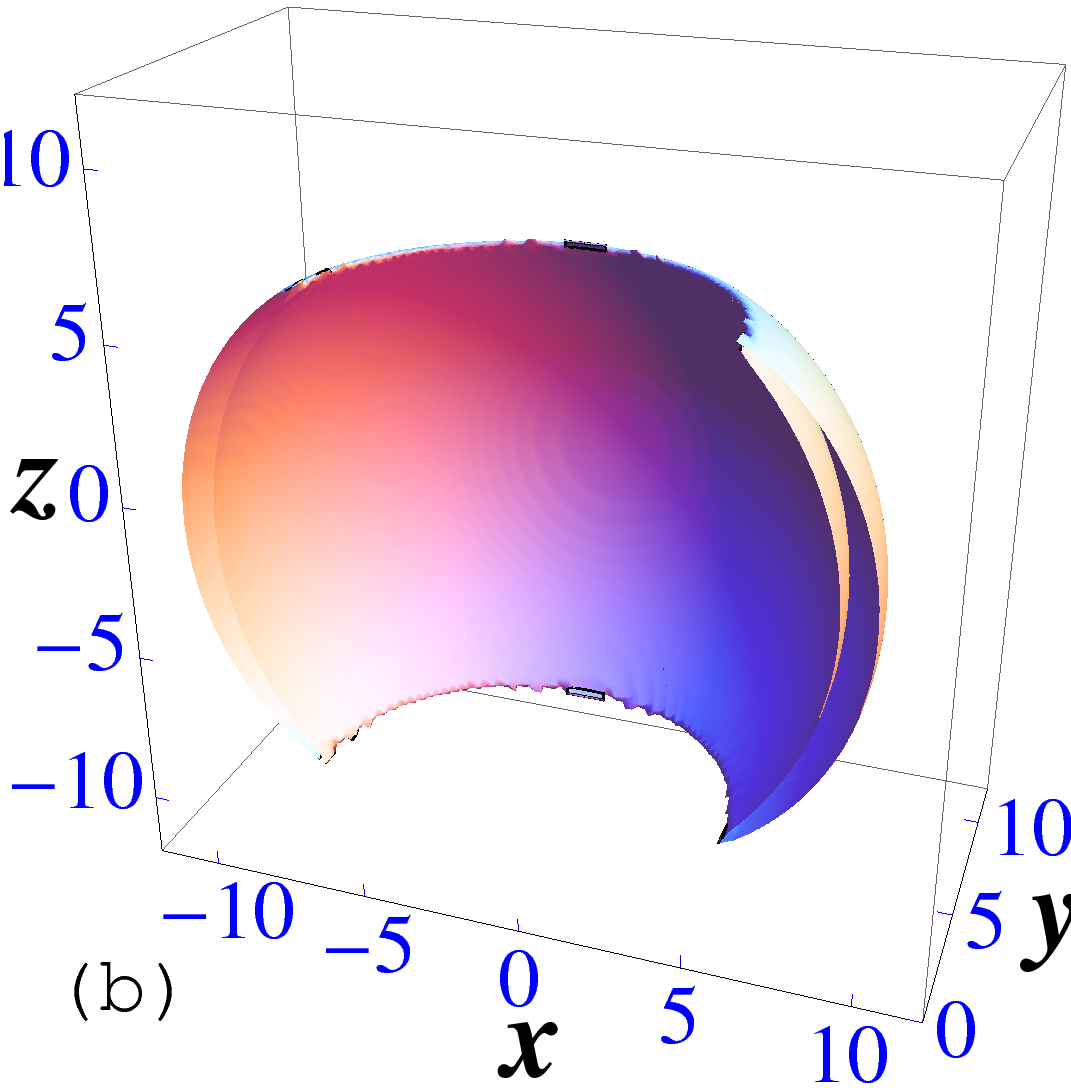}
\includegraphics[width=.49\linewidth,clip]{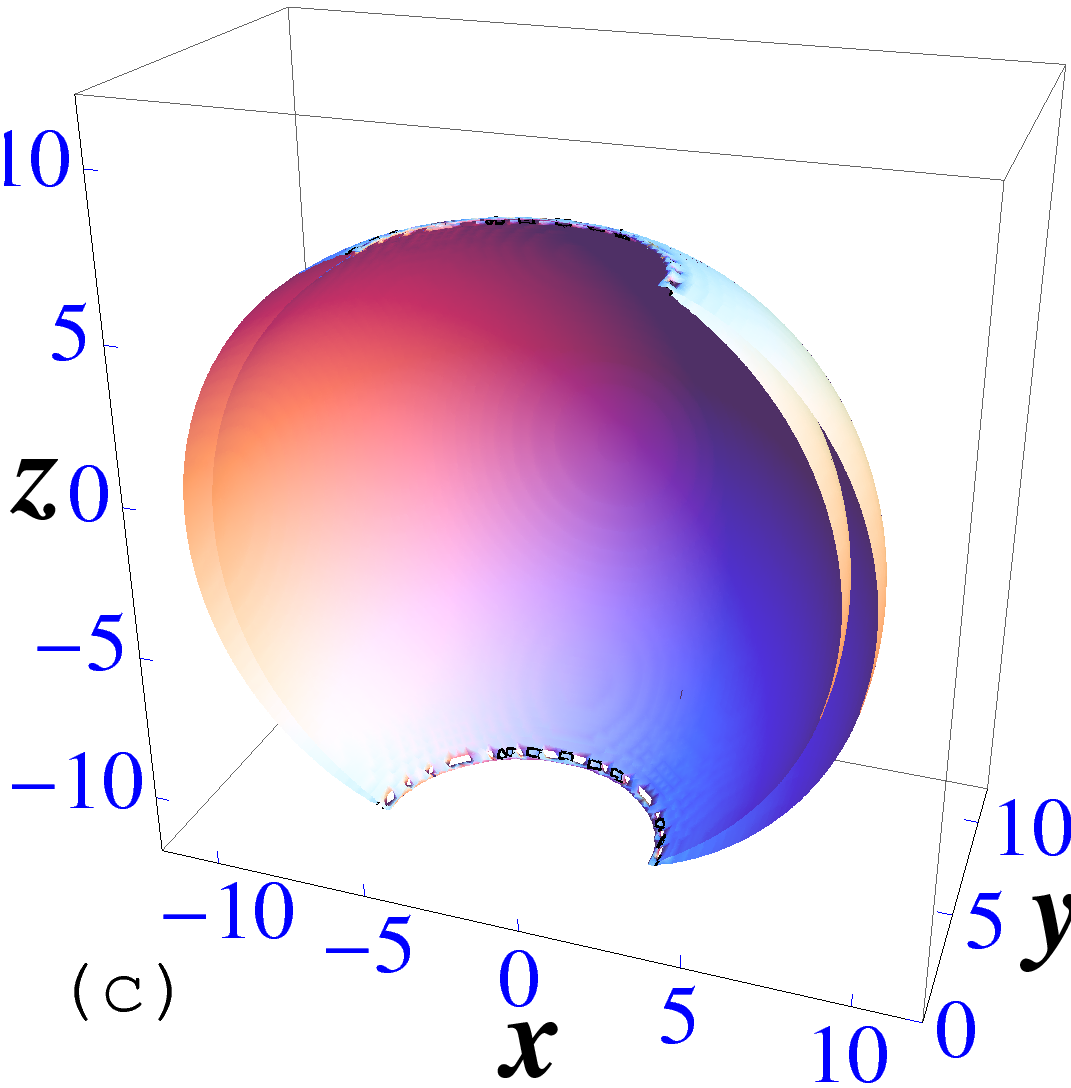}
\includegraphics[width=.49\linewidth,clip]{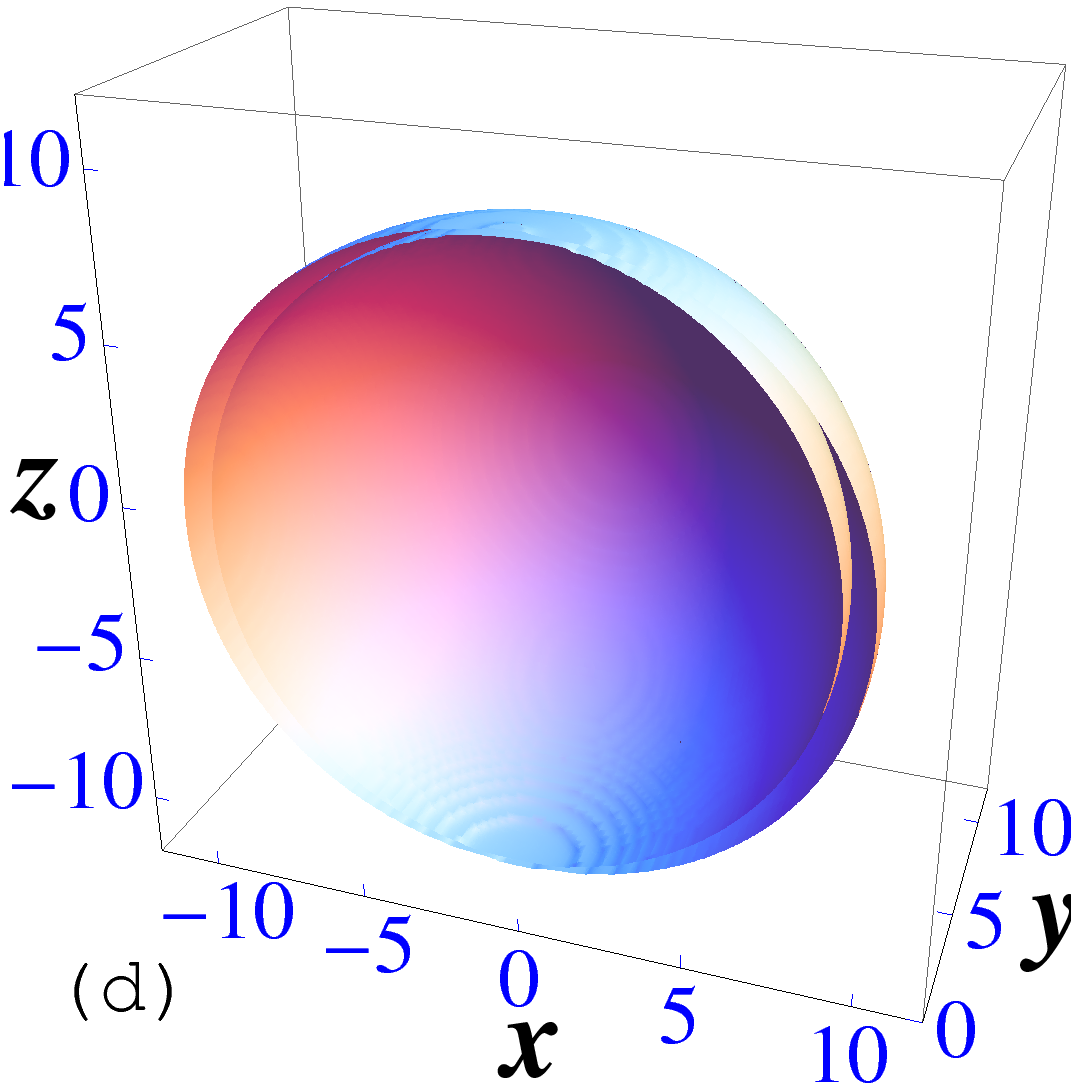}
\end{center}

\caption{(Color online) 
3D contour plot of density $|\phi|^2$  for a shell-shaped dipolar 
BEC with 
$ g_{dd}=100, 
$ and (a) $g=300$, (b)$g=600$, (c) $g=1000$, (d) $g=2000.$  
The density on the 
contour is 0.0005.
 {\color{red} The variables $x$, $y$, and $z$ are in units of $l_0
(= 1$ $\mu$m.)}
}  
\label{fig5}
\end{figure}

In the case of the ring the density is most asymmetric for $\alpha =\pi/2$ 
and we study the density in this case in some detail for a dipolar BEC 
of 5000 $^{52}$Cr atoms with $a_{dd}=15a_0$ for different values of contact interaction. 
For this purpose, we show the 2D contour plot of density in the $x-z$ plane 
$|\phi(x,y=0,z)|^2 $ in Figs. \ref{fig4} (a), (b), (c), and (d) for $a= 20a_0,
50a_0, 70a_0$ and $100a_0,$ respectively. The anisotropic dipolar interaction is most 
pronounced when the contact interaction is the smallest, e.g., for $a=20a_0$ 
in Fig. \ref{fig4} (a). In this case the anisotropic distribution of density 
in the plane of the ring is most visible. 
As the isotropic contact interaction increases, 
the effect of the anisotropic dipolar interaction 
is less and less pronounced and an almost symmetric density distribution 
in the plane of the ring is obtained for $a=100a_0$ as can be seen in 
Fig. \ref{fig4} (d).

In Figs. \ref{fig5} (a), (b), (c), and (d) we show the 3D contour plot of density $|\phi|^2$  
for $g_{dd}=100$ and different $g=300,600,1000$  and 2000.
For a clean visualization of the spherical-shell shape only one half of the full 
density distribution is shown. The density is most asymmetric in 
the polarization $z$ direction  in Fig. \ref{fig5} (a) with the smallest 
contact nonlinearity $g$, while the dipolar interaction is most prominent.
This asymmetry in density reduces as the 
contact nonlinearity increases making the anisotropic dipolar interaction 
less and less prominent. In Fig. \ref{fig5} (d), the density distribution is
the most isotropic in 3D with dipole interaction playing a minor role for 
$g_{dd}=100$ and $g=2000$.

\begin{figure}[!t]
\begin{center}
\includegraphics[width=\linewidth,clip]{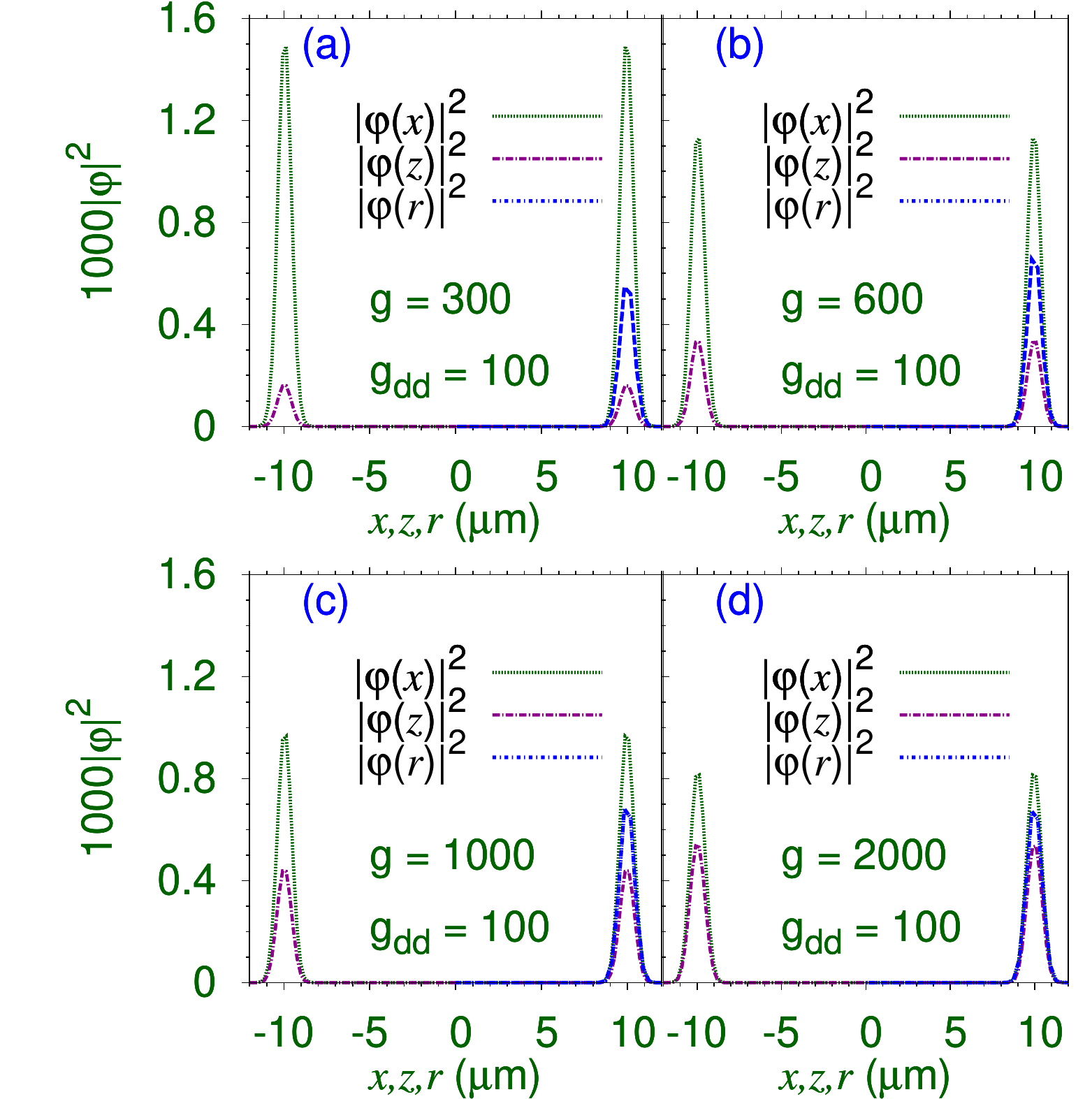}
\end{center}

\caption{(Color online) Densities $|\varphi(x)|^2 \equiv
|\phi(x,0,0)|^2, |\varphi(z)|^2 \equiv |\phi(0,0,z)|^2, |\varphi(r)|^2 \equiv 
|\phi(r,r,r)|^2,$ along $x$, $z$, and $r$ radial directions 
for a shell-shaped dipolar BEC with 
$g_{dd}=100$ and $g=$ (a) 300, (b) 600, (c) 1000, and (d) 2000. 
}  
\label{fig6}
\end{figure}

The above anisotropic density distribution of the shell-shaped dipolar BEC 
is further demonstrated by plotting the density along the $x$, $z$, and the 
radial $r$ directions in Figs. \ref{fig6} (a), (b), (c), and (d) for 
$g_{dd} =100$ and $g=300,600,1000$ and 2000, respectively. The anisotropy in the density in the three directions is explicitly shown in these 
figures. The anisotropy clearly reduces as the contact nonlinearity $g$ increases making the dipolar interaction less prominent, as can 
be seen in Fig. \ref{fig6} (d) with the most isotropic density distribution.

\subsection{Vortex dipolar BEC}

\begin{figure}[!b]
\begin{center}
\includegraphics[width=\linewidth,clip]{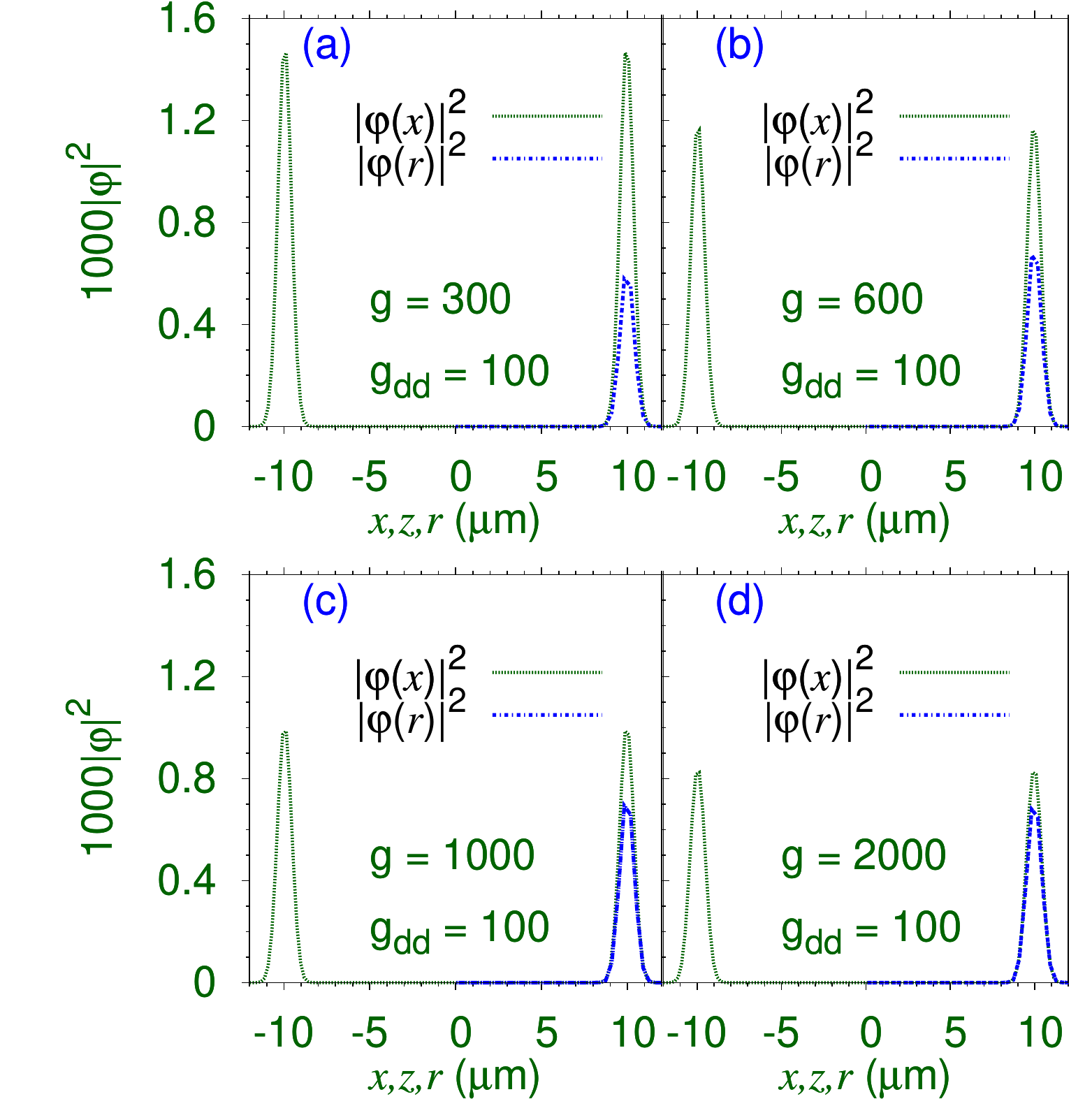}
\end{center}

\caption{(Color online) Densities $|\varphi(x)|^2 \equiv
|\phi(x,0,0)|^2$ and $|\varphi(r)|^2 \equiv 
|\phi(r,r,r)|^2$ along $x$ and $r$  directions 
for a shell-shaped vortex dipolar BEC with 
$g_{dd}=100$ and $g=$ (a) 300, (b) 600, (c) 1000, and (d) 2000. 
}  
\label{fig7}
\end{figure}

Now we study the singly-quantized vortices in ring- and shell-shaped 
dipolar BECs. In such a vortex, the trapping potential should have azimuthal 
symmetry. In the ring shape, such symmetry exists only for $\alpha =0$.  
However, for $\alpha =0$ the matter is localized far away from the center
[viz. Fig. \ref{fig3} (a)]. Consequently, the added centrifugal part in 
Eq. (\ref{potv}) is practically zero for large $x$ and $y$ ($\approx 10$) and 
the density of the 
vortex state is practically  identical with the normal state
shown in Fig. \ref{fig3} (a). In the $x-y$ plane, both have zero density at 
the center of the ring.  Nevertheless, the phase of ring vortex wave function
of small  finite radius 
changes by 2$\pi$ as one moves in a closed contour around the center of the 
vortex.
However, for larger values of angular momentum of the ring-shaped BEC
there could be dissipation when 
the rotational velocity exceeds a critical value \cite{rs2,x4},  
larger than the velocity for the unit angular momentum considered 
here. Below this
critical velocity the mass flow is dissipationless in analogy to
electrical current in superconductors. Such a persistent flow
was observed in a nondipolar BEC in an optically plugged
magnetic trap \cite{tor2}.

Next we consider 
 a   shell-shaped dipolar BEC rotating around the polarization $z$ direction 
with unit angular momentum. For this purpose, we solve the GP equation 
with potential (\ref{potv}).  For the vortex state the density on the $z$ axis 
$|\varphi(z)|^2$
becomes zero. Otherwise, the added centrifugal term in Eq. (\ref{potv})
is very small for large $x$ and $y$. For   most parts of the shell the 
values of $x$ and $y$ are large and the centrifugal part of the potential has 
very little effect on the density of the condensate. Consequently, the density 
distribution of the dipolar vortex BEC will be very similar to that of 
the normal BECs shown in Figs. \ref{fig6}, with the only different that for the vortex
the density 
at all points on the $z$ axis will be zero: $|\varphi(z)|^2=0$. This is illustrated in  Figs. \ref{fig7}
where we plot the densities $|\varphi(x)|^2$ and $|\varphi(r)|^2$
as in Figs. \ref{fig6}. It is seen that the densities in 
$x$ and $r$ directions of the normal and vortex BECs are quite the same.
As the density distributions of the normal 
and vortex BECs are similar, the stability phase plot of Fig. \ref{fig1} and the 
chemical potential plot of Fig. \ref{fig2} for the shell-shaped 
normal and vortex dipolar BEC are practically 
the same.

\begin{figure}[!t]
\begin{center}
\includegraphics[width=\linewidth,clip]{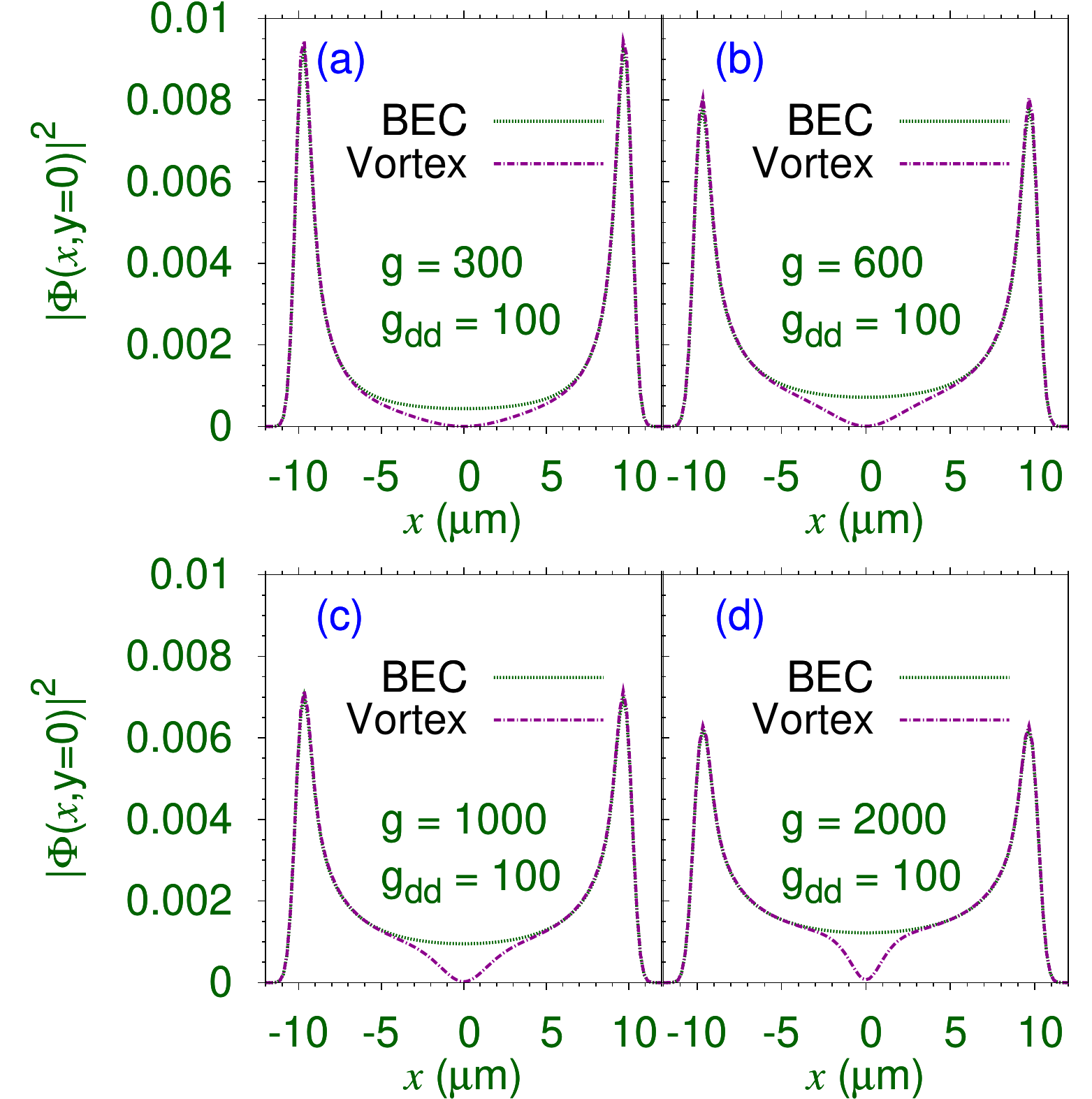}
\end{center}

\caption{(Color online)  Radial 2D density along $x$ axis $|\Phi(x,y=0)|^2=\int dz |\phi(x,y=0,z)|^2$ 
of a shell-shaped 
normal and vortex dipolar BEC  
with $g_{dd}=100$ and $g=$ (a) 300, (b) 600, (c) 1000, and (d) 2000. 
}  
\label{fig8}
\end{figure}

The difference in density distribution 
between the shell-shaped 
normal and the vortex BEC can be best studied by considering the isotropic 
two-dimensional (2D)
density of the condensate $|\Phi(x,y)|^2 = \int dz |\phi(x,y,z)|^2 $ obtained by 
integrating out the $z$ dependence. For a vortex this 2D density is zero for $x=y=0$. 
To illustrate the difference in density distribution of a normal and a vortex BEC, we plot 
in Fig. \ref{fig8} the density $ |\Phi(x,0)|^2$ versus $x$ for the 
different cases shown in Figs. \ref{fig6}. For these sets of parameters the densities for 
the normal and the vortex BECs are shown. The two densities are quite similar over most
regions except near the center $x=0$. For the vortex BEC the density is zero at the center, 
whereas for the normal BEC it has a small finite value.

\section{Conclusion}

We studied the properties of a normal (nonrotating) and vortex dipolar BEC in ring- and 
shell-shaped traps using the mean-field GP equation. The stability of the system with 
anisotropic density distribution, which is a consequence of the anisotropic dipolar 
interaction, is studied in phase plots of dipolar and contact interactions. The system is 
more stable with reduced anisotropic density distribution when  the dipolar interaction is 
small compared to the contact interaction. In the shell-shaped trap the dipolar BEC is 
always unstable when the contact interaction is zero or attractive. In the ring-shaped trap 
the dipolar BEC can be stable for zero contact interaction when the plane of the ring makes an
angle with the polarization direction ($\alpha \ne \pi/2$), as can be seen from the stability
plot in Fig. \ref{fig1}.

The vortex dipolar BECs are also 
considered in the axially-symmetric ring- and shell-shaped traps.  The ring-shaped trap is axially symmetric 
when the ring stays in a plane perpendicular to the polarization direction. In the case of the ring-shaped 
trap the density distribution of the normal and vortex dipolar BECs is practically the same. The difference in the 
density distribution  of the normal and vortex dipolar BECs in the shell-shaped trap is carefully examined. 
The experimental realization of the shell- and ring-shaped traps 
\cite{rs1,rs2}
and the present theoretical study will trigger further studies of dipolar 
BEC in these novel traps.

\acknowledgments
We thank
FAPESP and    CNPq (Brazil)
for partial support.


\end{document}